\documentclass[sigconf,review=false]{acmart}

\usepackage{listings}
\AtBeginDocument{%
  \providecommand\BibTeX{{%
    \normalfont B\kern-0.5em{\scshape i\kern-0.25em b}\kern-0.8em\TeX}}}

\setcopyright{acmcopyright}
\copyrightyear{2024}
\acmYear{2024}
\acmDOI{XXXXXXX.XXXXXXX}

\acmConference[ITiCSE '24]{Proceedings of the 2024 ACM Conference on Innovation and Technology in Computer Science Education}{July 8--10, 2024}{Milan, Lombardy, Italy}
\acmISBN{978-1-4503-XXXX-X/18/06}

\usepackage{geometry}
\begin{document}

\title{A Comprehensive Analysis of Student-Generated Analogies with Large Language Models in CS1}
\title{``Like a Nesting Doll’’: Analyzing Recursion Analogies Generated by CS Students using Large Language Models}
\author{Seth Bernstein}
\affiliation{
  \institution{Temple University}
  \city{Philadelphia}
  \state{PA}
  \country{United States}}
\email{seth.bernstein@temple.edu}

\author{Paul Denny}
\affiliation{%
  \institution{University of Auckland}
  \city{Auckland}
  \country{New Zealand}}
\email{paul@cs.auckland.ac.nz}

\author{Juho Leinonen}
\affiliation{%
  \institution{University of Auckland}
  \city{Auckland}
  \country{New Zealand}}
\email{juho.leinonen@auckland.ac.nz	}

\author{Lauren Kan}
\affiliation{%
  \institution{Temple University}
  \city{Philadelphia}
  \state{PA}
  \country{US}}
\email{lauren.kan@temple.edu}

\author{Arto Hellas}
\affiliation{
  \institution{Aalto University}
  \city{Espoo}
  \country{Finland}}
\email{arto.hellas@aalto.fi}

\author{Matt Littlefield}
\affiliation{%
  \institution{Temple University}
  \city{Philadelphia}
  \state{PA}
  \country{US}}
\email{matt.littlefield@temple.edu}

\author{Sami Sarsa}
\affiliation{
  \institution{Aalto University}
  \city{Espoo}
  \country{Finland}}
\email{sami.sarsa@aalto.fi}

\author{Stephen MacNeil}
\affiliation{%
  \institution{Temple University}
  \city{Philadelphia}
  \state{PA}
  \country{US}}
\email{stephen.macneil@temple.edu	}

\renewcommand{\shortauthors}{Anon, et al.}

\begin{abstract}


Grasping complex computing concepts often poses a challenge for students who struggle to anchor these new ideas to familiar experiences and understandings.
To help with this, a good analogy can bridge the gap between unfamiliar concepts and familiar ones, providing an engaging way to aid understanding.
However, creating effective educational analogies is difficult even for experienced instructors.  
We investigate to what extent large language models (LLMs), specifically ChatGPT, can provide access to personally relevant analogies on demand. Focusing on recursion, a challenging threshold concept, we analyzed the analogies generated by 385 first-year computing students. 
They were provided with a code snippet and tasked to generate their own recursion-based analogies using ChatGPT, optionally including personally relevant topics in their prompts. 
We observed a great deal of diversity in the analogies produced with student-prescribed topics, in contrast to the otherwise generic analogies, highlighting the value of student creativity when working with LLMs. 
Not only did students enjoy the activity and report an improved understanding of recursion, but they described more easily remembering analogies that were personally and culturally relevant. 

\end{abstract}





\maketitle

\section{Introduction}

Learning to program is challenging because students may struggle to connect unfamiliar terms and concepts to existing knowledge and to their everyday lived experiences. Especially for difficult threshold concepts~\cite{sanders2016threshold, boustedt2007threshold}, like recursion, students benefit from high-quality explanations or visual representations such as visualizations and concept maps~\cite{sanders2016threshold}. Analogies and metaphors are another useful pedagogical technique which connect unfamiliar concepts to ideas that students already intuitively understand~\cite{heywood2002place, coll2005role,gilbert2016analogies,seel2017model}. Analogies and metaphors have been previously studied in computing education contexts~\cite{forivsek2012metaphors, bettin2021frozen, bettin2022semaphore, larsson2023metaphors, cao2016examining, harper2022developing, bettin2023more, giacaman2012teaching}. However, it has been shown that students struggle to craft high-quality analogies and that they could benefit from additional scaffolding~\cite{harper2023investigating}. 

Large language models (LLMs) have recently been shown to generate high-quality explanations~\cite{macneil2023experiences, leinonen2023comparing}. Students even prefer explanations generated by an LLM to ones generated by their peers~\cite{leinonen2023comparing}. Consequently, LLMs may be useful tools for scaffolding students trying to craft analogies. However, 
although there is some evidence that LLMs can generate analogies~\cite{macneil2022generating, ding2023fluid},
there has not yet been any systematic investigation into how students might use LLMs to generate their own analogies. Given this gap in the literature, we investigated the following research questions: 


\begin{itemize}
    \item[\textbf{RQ1}] What types of analogies do students generate to explain recursion when given access to an LLM?
    \item[\textbf{RQ2}] What do students think about using LLMs to generate their own analogies?  
\end{itemize}
In this paper, we conducted a study with a sample size of over 350 students at a large public research university. As part of their lab on recursion, a well-known threshold concept that is challenging for many students~\cite{sanders2016threshold, boustedt2007threshold}, students were asked to use ChatGPT to craft their own analogies to explain recursion using any personally relevant topic as a theme. We collected student responses to survey questions and analyzed their generated analogies and corresponding prompts. Our results show that when students explicitly specified the topic of the analogy in their prompt, the resulting analogies were more diverse, featuring a different distribution of topics compared to those produced by the LLM when topics were not prescribed. This suggests that involving students in crafting analogies with ChatGPT leads to more creative and relevant analogies compared to relying solely on the model. Students expressed positive sentiments about the activity, noting instances where the model-generated analogies aided their understanding and retentionPrevious studies have established the effectiveness of analogies in computing education~\cite{bettin2021frozen, bettin2022semaphore, cao2016examining, harper2022developing, bettin2023more, giacaman2012teaching}, which is not the focus of this paper. Instead, this study focuses on what analogies students produce and their potential for engagement. Future work will focus on systematically investigating the impact on student learning.

\section{Related Work}

\subsection{Analogies to Explain Computing Concepts}

Explanations are critical for helping students understand complex topics. Over two decades of computing education research has demonstrated how practicing `explain in plain English' questions~\cite{whalley2006australasian} and self-explanations~\cite{vihavainen2015benefits, murphy2012ability, vieira2017writing} can provide students with short-term and long-term benefits. More recently, computing education researchers have begun investigating the use of analogies to make unfamiliar concepts more approachable for students. Often, analogies are carefully developed by instructors to convey complex concepts such as parallel programming~\cite{giacaman2012teaching, bettin2022semaphore, bettin2023more}, introductory computing~\cite{harper2023investigating, cao2016examining, heinonen2020exploring}, and algorithms and data structures~\cite{forivsek2012metaphors}. 

Analogies are a pedagogical tool that equates two disparate concepts based on their similarities in at least one aspect. In computing education, where many unfamiliar terms and concepts may seem completely disconnected from previously learned concepts and lived experiences, analogies have the ability to connect these complex concepts to ones that are more familiar to the learner. For example, an algorithm is like a recipe, a bitwise operator is like a light switch, and a memory address is like a house number. Analogies do not need to perfectly describe the concept to be useful. Analogies help to provide an intuition for the concept which may help students get passed an initial hurdle in understanding. In recent work, Harper et al. investigated the potential of student-generated analogies in computing education, particularly in fostering comprehension of abstract computing concepts like variable scope \cite{harper2023investigating}. Their work showed that students faced challenges in achieving complete and precise structural alignment, along with maintaining an appropriate level of abstraction across their analogy mappings. Bettin et al. showed that when providing students with analogies, preferences about the analogies varied across students and students did not always use a given analogy in the same way~\cite{bettin2022semaphore, bettin2023more}. These prior studies suggest that students benefit from analogies but that analogies are ideally personalized based on each students needs and experiences. Additionally, students could greatly benefit from targeted scaffolding to address these challenges and enhance their ability to generate effective analogies.




\subsection{Generating Learning Materials with LLMs}

Large language models (LLMs) provide opportunities for educators to generate many high-quality learning materials at scale~\cite{macneil2022automatically, finnie2022robots, macneil2022generating, macneil2023experiences, leinonen2023comparing}. Finnie-Ansley et al. demonstrated that Codex could produce programming assignments that were useful as a first draft for educators~\cite{finnie2022robots}. Since then, LLMs have been used to generate explanations~\cite{macneil2022generating, macneil2023experiences, leinonen2023comparing, leinonen2023using, macneil2024decoding}, programming assignments~\cite{sarsa2022automatic, tran2023using}, and to respond to students' help requests~\cite{hou2024effects, Hellas_2023}. More recent work has provided empirical evidence that students are legitimately relying on Generative AI as a source of help when they get stuck~\cite{prather2023robots, hou2024effects, cindyGenerating}. 

Several papers have recently explored the potential of LLMs to enhance code comprehension, with a particular focus on their ability explain code to students. In an initial study by MacNeil et al., the researchers demonstrated that LLMs were able to trace code execution, articulate the high-level purpose, and offer analogies for programming concepts~\cite{macneil2022generating}. This initial investigation focused on the capabilities but not on the associated quality of the explanations and analogies. MacNeil et al. later extended their investigating by studying students' preferences and engagement with explanations generated by LLMs, considering variations in the types of explanation provided~\cite{macneil2023experiences}. Leinonen et al. demonstrated that the explanations produced by LLMs were preferred by students when compared to explanations generated by their peers~\cite{leinonen2023comparing}. Despite the considerable attention LLM-generated explanations have received, there remains a gap in the literature regarding a systematic study of analogies generated by students with the assistance of LLMs.

\section{Method}

\subsection{Study Context}

The data for this study were collected from an introductory programming course taught in Spring 2023 at the University of Auckland, a large public research University in New Zealand. The 12-week course used C as the programming language and is a required course for students in the first year of the Engineering program. These students specialize in their second year based on GPA, and thus students in the course have an incentive to perform well in all of their first-year courses, even if they do not intend to specialize in computing topics. 

When our data were collected, 889 students were enrolled in the course and we collected 841 responses (95\% of enrolled students). The course featured weekly lab sessions that generally included programming exercises that are automatically graded. Ethics approval was granted by the institution (\emph{UAHPEC25279}).


\subsubsection{Recursion:}  The topic of recursion is introduced in the penultimate week of the course, and is the focus of the final lab in Week 12. As a threshold concept~\cite{sanders2016threshold}, recursion is considered a difficult topic for many students. This is supported by the literature~\cite{robins2003learning, gotschi2003mental}, as well as comments submitted by students on the end-of-semester teaching evaluations (examples from 2022 included: \emph{``spend some more time on recursion because I feel like it's quite a difficult topic to get on top of''} and \emph{``I found the visual representation of recursion and memory quite hard to get my head around''}). Furthermore, understanding recursion goes beyond memorizing definitions; it encompasses the ability to apply, test, and utilize the concept in real-world scenarios, thereby opening up new avenues for mental models and program design. The intricacies of recursion include both theoretical significance and substantial practical application, making it a transformative element of computing education \cite{rountree2013elaborating}.


\subsection{Materials}

In addition to the standard lecture materials used in the course for teaching recursion, which included recursive images (i.e. the Droste effect\footnote{\url{https://en.wikipedia.org/wiki/Droste_effect}}), visualizations of stack frames during recursive calls, and common but simplistic text-book examples (such as factorial and Fibonacci), students were also provided with an example of an analogy as part of the preparation materials for the Week 12 lab. This example was prefaced by the statement: \textit{``Analogies can be useful for teaching difficult concepts because they relate unfamiliar ideas to something already known, making the new information easier to understand. For instance, an analogy often used in physics textbooks compares electrical circuits to hydraulic circuits''}, and then demonstrated the output from ChatGPT when asked to \textit{``use an analogy to explain how electrical circuits work''.}  The output compared electrical circuits to a plumbing system, drawing comparisons between voltage/pressure, current/flow rate, and resistance/constriction.

The first exercise in the lab asked students to generate an analogy using ChatGPT to explain a recursive function that appeared in the course textbook.  The function prints a string in reverse, and is shown in Figure \ref{fig:print_reverse}.  The instructions given to students for this exercise are shown in Figure \ref{fig:analogy_instructions}.  We assumed that many (but not all) students had OpenAI accounts for ChatGPT access.  To ensure all students had access, we also provided a clone that we had built which served as an interface to ChatGPT but did not require a login. 

{
\begin{figure}[ht]
\small 
\centering
\begin{lstlisting}[language=C]
void PrintReverse(char *word)
{
   if (*word == '\0') {
      return;
   } else {
      PrintReverse(word+1);
      printf("%c", *word);   
   }
}
\end{lstlisting}
\caption{Function to print a string in reverse, for which students were asked to generate an analogy.}
\label{fig:print_reverse}
\end{figure}
}

\begin{figure}[ht]
\small
\centering
\fbox{%
    \begin{minipage}{\dimexpr\columnwidth-2\fboxsep-2\fboxrule\relax}

Try using ChatGPT to create a few analogies to help explain how the PrintReverse() function uses recursion to print a string in reverse.  Your first one might not be that good so try different prompts. Here are some tips:

    \textbf{Clarity.} The prompt should clearly define what you want the model to do. Avoid ambiguity. For instance, instead of "Tell me about dogs," use "Provide a detailed description of the characteristics, behavior, and care required for domestic dogs."
    
    \textbf{Context}. ChatGPT responds to the immediate context of the prompt. So, establishing a clear context is crucial. For example, the prompt "Translate the following English text to French: 'Hello, how are you?'" provides clear context and instructions.
    
    \textbf{Themes}. Analogies are most effective when they use themes that you are familiar with or interested in.  Consider choosing a theme (like sport, books, or anything else you are interested in) to help generate an interesting analogy.
    
    \textbf{Precision}. Precise prompts yield precise responses. For example, if you want a list, specifically ask for it: "List the top 10 most populous countries in the world."

Choose your favorite analogy and submit it in the space below, along with the prompt (you provided to ChatGPT) used to create it. Please also describe what you like about the analogy you created and why you think it would be helpful to understand the provided code.

    \end{minipage}%
}
\caption{Instructions provided to students for Exercise 1.}
\label{fig:analogy_instructions}
\end{figure}




\subsection{Data Collection and Analysis} 
The analogies that students created could connect to any topic or theme of their choosing (see the relevant instructions under `Themes' in Figure \ref{fig:analogy_instructions}). 
Based on the 841 responses, we observed 456 instances (54.2\%) where students did not include the prompt they used to generate the analogy, even though they were instructed to do so. That data was discarded from this analysis. 

\subsubsection{Stratifying the Analogies by Prompt Topic}

Using the remaining 385 analogies which contained an analogy and a prompt, we stratified the data based on whether the prompt explicitly specified a topic. This sample sized was used for all analysis, for example, one student used the prompt: \textit{``Can you make an analogy to do with Pokemon for this code: <code>''}. The student explicitly specified the topic (\textbf{Student-Selected}) to be `Pokemon'. An example where the student did not include a topic in their prompt is \textit{``provide me with a detailed and concise analogy of the code provided below to explain how the function uses recursion to print a string in reverse''}. In the absence of a student-prescribed topic, ChatGPT will produce an analogy using a topic it selects (\textbf{GPT-Selected}), which in this case of this example, was about a labyrinth.






\subsubsection{Identifying the Topics and Themes of the Analogies}

Each of the 385 analogies were coded based on their topic, with the prompts hidden to avoid bias. Each analogy could only be coded with one topic. Next, these topics were grouped thematically into 29 themes, with researchers mediating disagreements as needed. To enhance reliability, the researchers independently recoded all analogies using these themes. A subset of 150 randomly selected rows was used to compute inter-rater reliability using Cohen's Kappa, resulting in 0.83 for GPT-generated topics and 0.84 for student-generated analogies. Subsequent sections refer to these topics and themes. 


\subsubsection{Assessing the Diversity of the Analogy Topics}

We measured the diversity of analogy topics generated by ChatGPT by computing the number of distinct topics for the Student-Selected and GPT-Selected analogies. This number of distinct topics was divided by the number of total topics. This calculation resulted in a value between 0 and 1 where a higher number indicates greater diversity of topics across the set.




\subsubsection{Exemplar Analogies}

To exemplify the types of analogies produced by ChatGPT, we chose three exemplar analogies. Our selection process was subjective with the goal to identify analogies that are particularly engaging and insightful. This process involved reviewing a large randomly sampled subset of the analogies. We only reviewed analogies with associated student comments to report on students' perspectives about the chosen analogies. 

\subsubsection{Thematic Analysis of Students' Survey Responses} 

We analyzed students' open-responses using a reflexive thematic analysis method~\cite{braun2006using, braun2021one}. This process is `reflexive' because it acknowledges that the researcher plays an active role in identifying the themes. We followed the six phases which include familiarizing yourself with the data, generating initial codes, searching for themes, reviewing themes, and defining and naming the themes. At each phase, team members discussed the themes to mitigate interpretation biases.




\section{Results}

\subsection{RQ1: Analyzing the Generated Analogies} 

We found that only 226 prompts (63.1\%) contained an explicitly specified analogy topic (\textit{Student-Selected}). The average word count for analogies where the topic was not specified (\textit{GPT-Selected}) was 179.2 words. When students specified the topic, the average word count increased to 235.5 (a 31.4\% increase). This suggests that student-specified topics may yield more detailed analogies. In this section, we present examples of analogies generated by students along with their reflections on the analogies. We also compare and contrast analogies in terms of the topic distribution and diversity.


\begin{figure}[ht]
\footnotesize
\centering
\fbox{%
    \begin{minipage}{\dimexpr\columnwidth-2\fboxsep-2\fboxrule\relax}

Imagine you're a Pokémon Trainer on an adventure, and you've encountered a special ability called 'Reverse.' 
\\\\
You have a friendly Pikachu by your side. Now, you come across a series of Poké Balls with letters inside them instead of Pokémon.      As you walk through the tall grass, you find these letter-filled Poké Balls one by one. You pick up each Poké Ball, and Pikachu helps you examine its content. Instead of revealing the letters immediately, Pikachu tells you to keep it in your backpack for now.      You continue to explore, collecting more letter-filled Poké Balls as you go. After a while, you decide to take a break, and Pikachu suggests you start from the last Poké Ball you picked up. You retrieve the last Poké Ball from your backpack and examine it. Inside, you find a letter. You show it to Pikachu, and Pikachu says, 'This letter comes from the last Poké Ball we picked up. Let's keep it.' 
\\\\
You repeat this process, one Poké Ball at a time, starting from the last one you picked up and working your way back through your collection. Pikachu helps you reveal and remember the letters from each Poké Ball.      
\\\\
In the code, your 'PrintReverse' function is like Pikachu's assistance. It processes each character in the string (each Poké Ball), keeps it in memory, and then starts revealing the characters in reverse order, just like you go through your collection of Poké Balls in reverse order, revealing and remembering the letters one by one."   ====      PROMPT 2: Can you summarise that into steps?      

\begin{itemize}
    \item STEP 1: You are a Pokémon Trainer on an adventure with your Pikachu, and you come across letter-filled Poké Balls.  
    \item STEP 2: You collect these Poké Balls one by one and store them in your backpack without revealing the letters inside.      
    \item STEP 3: As you continue collecting, Pikachu suggests you start from the last Poké Ball you picked up.      
    \item STEP 4: You retrieve the last Poké Ball from your backpack, reveal a letter, and Pikachu notes that this letter comes from the last Poké Ball collected.      
    \item STEP 5: You repeat this process for each Poké Ball, always starting from the last one and working your way back through your collection.      
    \item STEP 6: Pikachu helps you reveal and remember the letters from each Poké Ball.  
\end{itemize}    

In the code, your 'PrintReverse' function follows a similar pattern. It processes each character in the string, keeps it in memory, and then starts revealing the characters in reverse order. Just like in the Pokémon analogy, you work your way backward through your collection of letters, revealing and remembering them one by one.
    \end{minipage}%
}
\caption {An Example Analogy Generated by P242}
\label{fig:analogy_example}
\end{figure}

\subsubsection{Example Analogies}

In this section, we present three personalized analogies generated by ChatGPT in response to prompts that specified a topic. Each analogy aligned with the student's interests and is accompanied by comments from the respective student.


\subsubsection*{Pokémon Trainer}
In this analogy, shown in Figure~\ref{fig:analogy_example}, the student imagines themselves as a Pokémon Trainer with a Pikachu, collecting letter-filled Poké Balls. The analogy provides a step-by-step process for `collecting' all of the letters and then printing them in reverse order. This scenario is analogous to manipulating a string array where each Poké Ball corresponds to a character variable in a string array. The student (P242) who generated this analogy commented about how LLMs give students the freedom to personalize the analogy topic, which increased their engagement: 

\begin{quote}
\textit{``I like it because its simple, and also fits within the Pokemon world (apart from the talking Pikachu part). Even though its not the most detailed, it gives you a good general idea of what it is. \textbf{The freedom of what theme of analogy you can ask... greatly increases the engagement/interest.}''} (P242)    
\end{quote}

\subsubsection*{Harry Potter and the Chamber of Secrets}
This analogy focused on the journey of Harry Potter through the Chamber of Secrets as an analogy for the traversing character pointers in code. The recursive exploration and revelation in the story are analogous to processing and reversing characters in a string. This student commented on the cultural relevance of this topic for them and other Gen Z students: 

\begin{quote}
\textit{``Using Harry Potter as an example for the analogy for Recursion fits well because \textbf{Harry Potter has a cult following and is widely popular amongst Gen Z}... it allows students to understand how recursion works given a familiar context... 
associating the code with a story like the Chamber of Secrets, the concept of recursion can provide a sense of engagement, which makes the recursion; a difficult concept can be grasped easier than other potential analogies.''
} (P277)
\end{quote}

\subsubsection*{Russian Nesting Dolls}
This last analogy was one of the most popular analogy topics. It uses the analogy of Russian nesting dolls to illustrate recursion. The base case is compared to the smallest doll, with each recursive call opening a larger doll, ultimately leading to the reversal of the order as the process completes. This student talked about how this connected to fond memories as a child and their personal connection to the topic: 
\begin{quote}
\textit{``The reason why I love this analogy so much is it use the concept of unwrapping Russian Doll, which was one of my favourite toys and cartoon when I was young. \textbf{In fact that when I heard \{Anonymized Instructor\} talked about recursion, I immediately remember the Russian Dolls.} The AI successfully utilise that concept and clearly explain how the code use recursion to work which make me understand it without much problem.''} (P163)
\end{quote}



\subsubsection{Comparing the Diversity and Distribution of Topics}

As shown in Table~\ref{tab:topics}, when students provided a specific topic for their prompts, the resulting analogies covered a much broader range of topics (0.747) than when no prompt topic was provided (0.284). Additionally, there were differences in theme popularity based on the presence or absence of a specified topic in the prompt. This suggests that the default topics that ChatGPT generates when not explicitly directed, do not fully align with the types of topics that students generate. This misalignment in interests suggests that there might be a disconnect between the natural inclinations of ChatGPT and the specific preferences or expectations of students. It raises questions about the effectiveness of default topic generation in meeting user needs, particularly in educational contexts where alignment with students' preferences aids meaningful engagement. Examples of particularly creative topics specified within students' prompts included these from P27, P136, P229, P511, and P18: 

\begin{itemize}
    \item \textit{``modern-day NSSL phase two launch vehicles''}  
    \item \textit{``a magic worm that eats the words''} 
    \item \textit{``Detective Recursive Rex''}
    \item \textit{``A grand book that contains the entire history of everything''}
    \item \textit{``Conveyor belt of sandwiches''}
\end{itemize}



\begin{table}[]
    \centering
    \caption{The top ten analogy themes and diversity of topics for analogies where the topic was specified in the analogy (student-generated) or not specified (GPT-generated). }
    \label{tab:topics}
    \begin{tabular}{c c | c c}
    \toprule
    \multicolumn{2}{c}{\textbf{Student-Generated}} & \multicolumn{2}{c}{\textbf{GPT-Generated}} \\
        \midrule
       Count & Theme & Count & Theme  \\
        \midrule
        27 & Food/Cooking & 53 & Russian Nesting Dolls  \\
        23 & Russian Nesting Dolls & 37 & Food/Cooking \\
        14 & Books & 31 & Books \\
        13 & Sports & 22 & Cards \\
        9 & Miscellaneous & 9 & Miscellaneous \\
        6 & Animals & 5 & Dominoes \\
        4 & Blocks/Legos & 4 & Mazes \\
        3 & Video Games & 3 & Transportation \\
        3 & Dominoes & 2 & Sports \\
        3 & Board games & 2 & Animals \\
        \midrule
        \multicolumn{2}{c|}{ \textit{Distinct Topics: 169}  } & \multicolumn{2}{c}{ \textit{Distinct Topics: 48} } \\ 
        \multicolumn{2}{c|}{ \textit{Total Topics: 226}  } & \multicolumn{2}{c}{ \textit{Total Topics: 169} } \\ 

        \multicolumn{2}{c|}{ \textit{Diversity of Topics: 0.747}  } & \multicolumn{2}{c}{ \textit{Diversity of Topics: 0.284} } \\ 
        \bottomrule
    \end{tabular}
\end{table}





\subsection{RQ2: Analyzing Student Reflections}

To understand students' experiences with the analogy generation activity, we conducted a thematic analysis of post-lab survey responses about the learning activity along with comments about the analogies that students generated. The analysis revealed key themes related to educational value and personal relevance. 

\subsubsection{Educational Value} When learning complex concepts like recursion, students can often encounter a significant barrier. Despite lectures and coding assignments designed to aid comprehension, many students described a crucial turning point in their understanding frequently emerged through the use of analogies. As P705 describes it, they did not get a `real' understanding of recursion until after the analogies:  
\begin{quote}
\textit{``It was initially difficult to wrap my head around recursive functions, and initially, the code I submitted was actually not recursive. \textbf{It was mostly after the analogies that I got a real understanding of what recursive functions were.} I feel, that the concepts could be explained a bit more in the lectures.
''} 
\end{quote}

This feedback is further exemplified by another student's reflection where they detailed the limited effectiveness of coding tasks alone in grasping complex concepts like recursion. In their experience, simply following procedural instructions was not enough to foster a deep understanding. The student's revelation occurred while engaging with a multiple analogies:

\begin{quote}
\textit{``I did the coding tasks first, and I didn't really understand recursion. I was just simply following the preparation document (which was very helpful for this lab). However, doing the analogy section \textbf{helped me understand recursion a lot better after reading so many analogies lol.}
''} (P584)
\end{quote}

The emphasis on `so many' analogies reflects the extensive exploration required to grasp a challenging concept, like recursion, and it also implies that it may be valuable to engage with multiple analogies. 
These responses highlighted the role that analogies can play in complementing traditional teaching methods. They suggest that the analogies filled students' knowledge gaps which might not have been fully addressed through conventional teaching methods.

\subsubsection{Personal Relevance} Many students described iterating to find the right analogy. For example, 515 tried creating multiple: 


\begin{quote}\textit{``I tried creating few analogies based on different topics (eg. photo and school) before creating this analogy about dogs. However, \textbf{I think this analogy based on dogs was most interesting compared to other analogies} and if this analogy is provided to a person who is interested in pets, this analogy will be a very effective analogy for them to understand how the `PrintReverse' function works. Moreover, this analogy thoroughly explains how recursion works in this function as well.''}\end{quote}


Other students shared this opinion.
For example, P310 described struggling with an initial analogy because they did not have enough familiarity, but later found an analogy that resonated with them:

\begin{quote}
\textit{``I found this particular response from [ChatGPT] to be the most effective at helping me visualise the recursion process. \textbf{I initially asked it to provide a response on the topic of sports but its analogy, based around relay running, was fairly confusing. Hence, I asked [ChatGPT] to provide a response based on the theme of history.} I found this response to be more engaging as the reference to scrolls helped me further understand how recursion 'unravels' itself once the base case is reached.''} (P310)
\end{quote}


This ability to rapidly create and engage with analogies is a promising benefit of LLMs. Many students described generating multiple analogies and talked about how some made sense to them while others did not. 

\subsubsection{Overall Student Sentiment}

Overall, students were extremely positive about the activity. Based on manually coding the sentiment of 150 randomly selected student reflections about the analogies, we observed only 8 instances (5.3\%) contained a negative sentiment, with 7 of those instances also containing positive sentiments. P24 is the only instance that contained purely negative sentiment:  

\begin{quote}
    \textit{``This response took quite a number of attempts to produce, because to start \textbf{the responses were much too long. This one seems to have enough detail to get an understanding of the topic, but not too much that it becomes boring.}''}
\end{quote}

However, with over 90\% of the responses containing positive sentiments, it is clear that the analogies were useful for students. In fact, students even described wanting to continue to use ChatGPT to generate analogies in the future. For example, P57 said: 

\begin{quote}
\textit{``Very nice lab on recursion, also the idea of being able to use chatGPT to create analogies is quite a nice thing to think about and \textbf{I'll definitely will be using this analogy part of chat in the future.}''}   
\end{quote}


\section{Discussion}


In this study, over 800 students were asked to generate their own analogies as part of an introductory programming lab focusing on recursion. We observed that only 63.1\% of the students explicitly specified a topic when prompting ChatGPT. This provided us with the opportunity to stratify the data by the analogies where the topic was explicitly chosen by students and analogies with a topic chosen implicitly by ChatGPT. Without stratifying the data, it would be unclear if students chose the topic. We observed that students tended to come up with more diverse analogy topics on average and the prevalence of topics also varied between the two groups. This disconnect in topical prevalence and diversity suggests that students may benefit from being `in-the-loop' when generating analogies. This aligns with prior work~\cite{harrison2006teaching, bettin2021frozen}; specifically that analogies should be `dialectic, not didactic' where ``analogy use and creation should be guided and discussed, not simply `given' ''~\cite{bettin2021frozen}. So while LLMs can easily generate bespoke analogies at any time on-demand, the interests, creativity, and culture of students is necessary to guide the process. In this way, LLMs may provide new opportunities for students to engage with computing concepts in ways that are culturally relevant~\cite{morales2019computing, madkins2019culturally, franklin2020scratch}. Multiple students commented on how the analogies could be relevant for their generation, culture, or lived experiences. This new capability for students to generate personalized analogies with very little effort undermines debates about which analogy is `best' for teaching a specific concept~\cite{bettin2021frozen}. Theoretically, students could continue to generate analogies until they find one that resonates with them and facilitates their understanding. However, this may add fuel to debates about whether a weak analogy can be misleading or confusing for students. Prior work has shown that even poor analogies do not do ``more damage'' than not having an analogy~\cite{san1993applying, bettin2021frozen}.

Despite overwhelming positive comments and the consistent advantages highlighted by students, the analysis revealed a surprising trend. Contrary to our initial expectations, a non-negligible number of the analogies provided by students were rather conventional, centering around familiar concepts like books and food. This raises intriguing questions about the prevalence of certain archetypal analogies in educational settings. One surprising aspect was the limited incorporation of analogies related to individualized interests or niche hobbies, such as a specific musician's lyrical style or references to movies. Further investigation is warranted to understand this trend better. Potential explanations include students' reluctance to share personally relevant analogies, the perception that niche topics may be more engaging but less informative, or the possibility that their best analogies, which may not have been the first or the most ``personally relevant'', were not initially shared. Additionally, analogies are memorable, which is great, but if they're inaccurate, this may lead to \textbf{memorable misconceptions.} Future research should assess analogy accuracy, focusing on its impact on student trust and the retention of knowledge and misconceptions.
These findings raise questions about how much scaffolding students should have in crafting analogies. Should teachers encourage students to generate personally relevant analogies or provide ready-made prompts that allow students to insert a topic.








\subsection{Limitations}

This study encountered several limitations. Firstly, a large subset of students did not include their prompts along with their analogies, resulting in some data that was not able to be analyzed. While the large sample size helps mitigate these effects, it remains uncertain whether some students faced challenges in generating analogies, and this aspect cannot be examined without the complete dataset. Another limitation is the absence of information on how many analogies students produced before selecting their final one. Multiple students explicitly shared in their reflections that they created analogies for multiple topics. Therefore, the chosen analogy may not fully represent this diversity. Additionally, since students performed this activity as part of a class assignment, the nature of the topics selected might differ if students were generating analogies spontaneously for immediate learning support. Lastly, to streamline the study, we provided students with a single programming problem. It is unclear whether these findings would be replicated with a different problem, as topic representation may vary.

\section{Conclusion}


Our exploratory study investigated the types of analogies that students generated when using large language models (LLMs). Over 350 students generated analogies for a recursive code snippet and provided the corresponding prompt. One consistent finding was that students were largely positive about the learning activity. Students described the value they received regarding the ability to personalize the analogies. Participants described being able to easily remember analogies that were personally and culturally relevant. They also talked about how seeing multiple analogies provided additional value. Despite these benefits, we observed that only half of the students in this study took advantage of being able to personalize the analogies by providing an explicit topic in the prompt. It is possible that some students found sufficient value associated with generic analogies, but more work is need to understand why so many students failed to specify the analogy topic.




\bibliographystyle{ACM-Reference-Format}
\bibliography{references}


\begin{thebibliography}{44}


\ifx \showCODEN    \undefined \def \showCODEN     #1{\unskip}     \fi
\ifx \showDOI      \undefined \def \showDOI       #1{#1}\fi
\ifx \showISBNx    \undefined \def \showISBNx     #1{\unskip}     \fi
\ifx \showISBNxiii \undefined \def \showISBNxiii  #1{\unskip}     \fi
\ifx \showISSN     \undefined \def \showISSN      #1{\unskip}     \fi
\ifx \showLCCN     \undefined \def \showLCCN      #1{\unskip}     \fi
\ifx \shownote     \undefined \def \shownote      #1{#1}          \fi
\ifx \showarticletitle \undefined \def \showarticletitle #1{#1}   \fi
\ifx \showURL      \undefined \def \showURL       {\relax}        \fi
\providecommand\bibfield[2]{#2}
\providecommand\bibinfo[2]{#2}
\providecommand\natexlab[1]{#1}
\providecommand\showeprint[2][]{arXiv:#2}

\bibitem[Bettin and Ott(2021)]%
        {bettin2021frozen}
\bibfield{author}{\bibinfo{person}{Briana Bettin} {and} \bibinfo{person}{Linda Ott}.} \bibinfo{year}{2021}\natexlab{}.
\newblock \showarticletitle{Frozen in the Past: When it Comes to Analogy Fears, It's Time For Us to" Let it Go"}. In \bibinfo{booktitle}{\emph{Proceedings of the 26th ACM Conference on Innovation and Technology in Computer Science Education V. 1}}. \bibinfo{pages}{359--365}.
\newblock


\bibitem[Bettin et~al\mbox{.}(2022)]%
        {bettin2022semaphore}
\bibfield{author}{\bibinfo{person}{Briana Bettin}, \bibinfo{person}{Linda Ott}, {and} \bibinfo{person}{Julia Hiebel}.} \bibinfo{year}{2022}\natexlab{}.
\newblock \showarticletitle{Semaphore or Metaphor? Exploring Concurrent Students' Conceptions of and with Analogy}. In \bibinfo{booktitle}{\emph{Proceedings of the 27th ACM Conference on on Innovation and Technology in Computer Science Education Vol. 1}}. \bibinfo{pages}{200--206}.
\newblock


\bibitem[Bettin et~al\mbox{.}(2023)]%
        {bettin2023more}
\bibfield{author}{\bibinfo{person}{Briana Bettin}, \bibinfo{person}{Linda Ott}, {and} \bibinfo{person}{Julia Hiebel}.} \bibinfo{year}{2023}\natexlab{}.
\newblock \showarticletitle{More (Sema| Meta) phors: Additional Perspectives on Analogy Use from Concurrent Programming Students}. In \bibinfo{booktitle}{\emph{Proceedings of the 2023 Conference on Innovation and Technology in Computer Science Education V. 1}}. \bibinfo{pages}{166--172}.
\newblock


\bibitem[Boustedt et~al\mbox{.}(2007)]%
        {boustedt2007threshold}
\bibfield{author}{\bibinfo{person}{Jonas Boustedt}, \bibinfo{person}{Anna Eckerdal}, \bibinfo{person}{Robert McCartney}, \bibinfo{person}{Jan~Erik Mostr{\"o}m}, \bibinfo{person}{Mark Ratcliffe}, \bibinfo{person}{Kate Sanders}, {and} \bibinfo{person}{Carol Zander}.} \bibinfo{year}{2007}\natexlab{}.
\newblock \showarticletitle{Threshold concepts in computer science: do they exist and are they useful?}
\newblock \bibinfo{journal}{\emph{ACM Sigcse Bulletin}} \bibinfo{volume}{39}, \bibinfo{number}{1} (\bibinfo{year}{2007}).
\newblock


\bibitem[Braun and Clarke(2006)]%
        {braun2006using}
\bibfield{author}{\bibinfo{person}{Virginia Braun} {and} \bibinfo{person}{Victoria Clarke}.} \bibinfo{year}{2006}\natexlab{}.
\newblock \showarticletitle{Using thematic analysis in psychology}.
\newblock \bibinfo{journal}{\emph{Qualitative research in psychology}} \bibinfo{volume}{3}, \bibinfo{number}{2} (\bibinfo{year}{2006}), \bibinfo{pages}{77--101}.
\newblock


\bibitem[Braun and Clarke(2021)]%
        {braun2021one}
\bibfield{author}{\bibinfo{person}{Virginia Braun} {and} \bibinfo{person}{Victoria Clarke}.} \bibinfo{year}{2021}\natexlab{}.
\newblock \showarticletitle{One size fits all? What counts as quality practice in (reflexive) thematic analysis?}
\newblock \bibinfo{journal}{\emph{Qualitative research in psychology}} \bibinfo{volume}{18}, \bibinfo{number}{3} (\bibinfo{year}{2021}), \bibinfo{pages}{328--352}.
\newblock


\bibitem[Cao et~al\mbox{.}(2016)]%
        {cao2016examining}
\bibfield{author}{\bibinfo{person}{Yingjun Cao}, \bibinfo{person}{Leo Porter}, {and} \bibinfo{person}{Daniel Zingaro}.} \bibinfo{year}{2016}\natexlab{}.
\newblock \showarticletitle{Examining the value of analogies in introductory computing}. In \bibinfo{booktitle}{\emph{Proceedings of the 2016 ACM Conference on International computing education research}}. \bibinfo{pages}{231--239}.
\newblock


\bibitem[Coll et~al\mbox{.}(2005)]%
        {coll2005role}
\bibfield{author}{\bibinfo{person}{Richard~K Coll}, \bibinfo{person}{Bev France}, {and} \bibinfo{person}{Ian Taylor}.} \bibinfo{year}{2005}\natexlab{}.
\newblock \showarticletitle{The role of models/and analogies in science education: implications from research}.
\newblock \bibinfo{journal}{\emph{International Journal of Science Education}} \bibinfo{volume}{27}, \bibinfo{number}{2} (\bibinfo{year}{2005}), \bibinfo{pages}{183--198}.
\newblock


\bibitem[Ding et~al\mbox{.}(2023)]%
        {ding2023fluid}
\bibfield{author}{\bibinfo{person}{Zijian Ding}, \bibinfo{person}{Arvind Srinivasan}, \bibinfo{person}{Stephen MacNeil}, {and} \bibinfo{person}{Joel Chan}.} \bibinfo{year}{2023}\natexlab{}.
\newblock \showarticletitle{Fluid transformers and creative analogies: Exploring large language models' capacity for augmenting cross-domain analogical creativity}. In \bibinfo{booktitle}{\emph{Proceedings of the 15th Conference on Creativity and Cognition}}. \bibinfo{pages}{489--505}.
\newblock


\bibitem[Finnie-Ansley et~al\mbox{.}(2022)]%
        {finnie2022robots}
\bibfield{author}{\bibinfo{person}{James Finnie-Ansley}, \bibinfo{person}{Paul Denny}, \bibinfo{person}{Brett~A Becker}, \bibinfo{person}{Andrew Luxton-Reilly}, {and} \bibinfo{person}{James Prather}.} \bibinfo{year}{2022}\natexlab{}.
\newblock \showarticletitle{The Robots Are Coming: Exploring the Implications of OpenAI Codex on Introductory Programming}. In \bibinfo{booktitle}{\emph{Australasian Computing Education Conf.}} \bibinfo{pages}{10--19}.
\newblock


\bibitem[Fori{\v{s}}ek and Steinov{\'a}(2012)]%
        {forivsek2012metaphors}
\bibfield{author}{\bibinfo{person}{Michal Fori{\v{s}}ek} {and} \bibinfo{person}{Monika Steinov{\'a}}.} \bibinfo{year}{2012}\natexlab{}.
\newblock \showarticletitle{Metaphors and analogies for teaching algorithms}. In \bibinfo{booktitle}{\emph{Proceedings of the 43rd ACM technical symposium on Computer Science Education}}. \bibinfo{pages}{15--20}.
\newblock


\bibitem[Franklin et~al\mbox{.}(2020)]%
        {franklin2020scratch}
\bibfield{author}{\bibinfo{person}{Diana Franklin}, \bibinfo{person}{David Weintrop}, \bibinfo{person}{Jennifer Palmer}, \bibinfo{person}{Merijke Coenraad}, \bibinfo{person}{Melissa Cobian}, \bibinfo{person}{Kristan Beck}, \bibinfo{person}{Andrew Rasmussen}, \bibinfo{person}{Sue Krause}, \bibinfo{person}{Max White}, \bibinfo{person}{Marco Anaya}, {et~al\mbox{.}}} \bibinfo{year}{2020}\natexlab{}.
\newblock \showarticletitle{Scratch Encore: The design and pilot of a culturally-relevant intermediate Scratch curriculum}. In \bibinfo{booktitle}{\emph{Proceedings of the 51st ACM technical symposium on computer science education}}. \bibinfo{pages}{794--800}.
\newblock


\bibitem[Giacaman(2012)]%
        {giacaman2012teaching}
\bibfield{author}{\bibinfo{person}{Nasser Giacaman}.} \bibinfo{year}{2012}\natexlab{}.
\newblock \showarticletitle{Teaching by example: using analogies and live coding demonstrations to teach parallel computing concepts to undergraduate students}. In \bibinfo{booktitle}{\emph{2012 IEEE 26th International Parallel and Distributed Processing Symposium Workshops \& PhD Forum}}. IEEE, \bibinfo{pages}{1295--1298}.
\newblock


\bibitem[Gilbert and Justi(2016)]%
        {gilbert2016analogies}
\bibfield{author}{\bibinfo{person}{John~K Gilbert} {and} \bibinfo{person}{Ros{\'a}ria Justi}.} \bibinfo{year}{2016}\natexlab{}.
\newblock \showarticletitle{Analogies in modelling-based teaching and learning}.
\newblock \bibinfo{journal}{\emph{Modelling-based teaching in science education}} (\bibinfo{year}{2016}), \bibinfo{pages}{149--169}.
\newblock


\bibitem[G\"{o}tschi et~al\mbox{.}(2003)]%
        {gotschi2003mental}
\bibfield{author}{\bibinfo{person}{Tina G\"{o}tschi}, \bibinfo{person}{Ian Sanders}, {and} \bibinfo{person}{Vashti Galpin}.} \bibinfo{year}{2003}\natexlab{}.
\newblock \showarticletitle{Mental Models of Recursion}. In \bibinfo{booktitle}{\emph{Proceedings of the 34th SIGCSE Technical Symposium on Computer Science Education}} \emph{(\bibinfo{series}{SIGCSE '03})}. \bibinfo{publisher}{Association for Computing Machinery}, \bibinfo{address}{New York, NY, USA}, \bibinfo{pages}{346–350}.
\newblock
\showISBNx{158113648X}


\bibitem[Harper(2022)]%
        {harper2022developing}
\bibfield{author}{\bibinfo{person}{Colton Harper}.} \bibinfo{year}{2022}\natexlab{}.
\newblock \showarticletitle{Developing and Evaluating Scaffolding for Student-Generated Analogies in CS1}. In \bibinfo{booktitle}{\emph{Proceedings of the 27th ACM Conference on on Innovation and Technology in Computer Science Education Vol. 2}}. \bibinfo{pages}{656--657}.
\newblock


\bibitem[Harper et~al\mbox{.}(2023)]%
        {harper2023investigating}
\bibfield{author}{\bibinfo{person}{Colton Harper}, \bibinfo{person}{Ryan Bockmon}, {and} \bibinfo{person}{Stephen Cooper}.} \bibinfo{year}{2023}\natexlab{}.
\newblock \showarticletitle{Investigating Themes of Student-Generated Analogies}. In \bibinfo{booktitle}{\emph{Proceedings of the ACM Conference on Global Computing Education Vol 1}} \emph{(\bibinfo{series}{CompEd 2023})}. \bibinfo{publisher}{Association for Computing Machinery}, \bibinfo{pages}{64–70}.
\newblock
\showISBNx{9798400700484}


\bibitem[Harrison and Treagust(2006)]%
        {harrison2006teaching}
\bibfield{author}{\bibinfo{person}{Allan~G Harrison} {and} \bibinfo{person}{David~F Treagust}.} \bibinfo{year}{2006}\natexlab{}.
\newblock \showarticletitle{Teaching and learning with analogies: Friend or foe?}
\newblock \bibinfo{journal}{\emph{Metaphor and analogy in science education}} (\bibinfo{year}{2006}).
\newblock


\bibitem[Heinonen and Hellas(2020)]%
        {heinonen2020exploring}
\bibfield{author}{\bibinfo{person}{Ava Heinonen} {and} \bibinfo{person}{Arto Hellas}.} \bibinfo{year}{2020}\natexlab{}.
\newblock \showarticletitle{Exploring the instructional efficiency of representation and engagement in online learning materials}. In \bibinfo{booktitle}{\emph{United Kingdom \& Ireland Computing Education Research conference.}} \bibinfo{pages}{38--44}.
\newblock


\bibitem[Hellas et~al\mbox{.}(2023)]%
        {Hellas_2023}
\bibfield{author}{\bibinfo{person}{Arto Hellas}, \bibinfo{person}{Juho Leinonen}, \bibinfo{person}{Sami Sarsa}, \bibinfo{person}{Charles Koutcheme}, \bibinfo{person}{Lilja Kujanpää}, {and} \bibinfo{person}{Juha Sorva}.} \bibinfo{year}{2023}\natexlab{}.
\newblock \showarticletitle{Exploring the Responses of Large Language Models to Beginner Programmers' Help Requests}. In \bibinfo{booktitle}{\emph{Proceedings of the 2023 {ACM} Conference on International Computing Education Research V.1}}. \bibinfo{publisher}{{ACM}}.
\newblock


\bibitem[Heywood(2002)]%
        {heywood2002place}
\bibfield{author}{\bibinfo{person}{Dave Heywood}.} \bibinfo{year}{2002}\natexlab{}.
\newblock \showarticletitle{The place of analogies in science education}.
\newblock \bibinfo{journal}{\emph{Cambridge Journal of Education}} \bibinfo{volume}{32}, \bibinfo{number}{2} (\bibinfo{year}{2002}), \bibinfo{pages}{233--247}.
\newblock


\bibitem[Hou et~al\mbox{.}(2024)]%
        {hou2024effects}
\bibfield{author}{\bibinfo{person}{Irene Hou}, \bibinfo{person}{Sophia Mettille}, \bibinfo{person}{Owen Man}, \bibinfo{person}{Zhuo Li}, \bibinfo{person}{Cynthia Zastudil}, {and} \bibinfo{person}{Stephen MacNeil}.} \bibinfo{year}{2024}\natexlab{}.
\newblock \showarticletitle{The Effects of Generative AI on Computing Students’ Help-Seeking Preferences}. In \bibinfo{booktitle}{\emph{Proceedings of the 26th Australasian Computing Education Conference}} \emph{(\bibinfo{series}{ACE '24})}. \bibinfo{publisher}{Association for Computing Machinery}, \bibinfo{address}{New York, NY, USA}, \bibinfo{pages}{39–48}.
\newblock
\showISBNx{9798400716195}
\urldef\tempurl%
\url{https://doi-org.libproxy.temple.edu/10.1145/3636243.3636248}
\showURL{%
\tempurl}


\bibitem[Janet~Rountree and Rountree(2013)]%
        {rountree2013elaborating}
\bibfield{author}{\bibinfo{person}{Anthony~Robins Janet~Rountree} {and} \bibinfo{person}{Nathan Rountree}.} \bibinfo{year}{2013}\natexlab{}.
\newblock \showarticletitle{Elaborating on threshold concepts}.
\newblock \bibinfo{journal}{\emph{Computer Science Education}} \bibinfo{volume}{23}, \bibinfo{number}{3} (\bibinfo{year}{2013}), \bibinfo{pages}{265--289}.
\newblock


\bibitem[Larsson(2023)]%
        {larsson2023metaphors}
\bibfield{author}{\bibinfo{person}{Andreas Larsson}.} \bibinfo{year}{2023}\natexlab{}.
\newblock \showarticletitle{Metaphors and Gestures in Programming Education}.
\newblock In \bibinfo{booktitle}{\emph{Programming and Computational Thinking in Technology Education}}. \bibinfo{publisher}{Brill}.
\newblock


\bibitem[Leinonen et~al\mbox{.}(2023a)]%
        {leinonen2023comparing}
\bibfield{author}{\bibinfo{person}{Juho Leinonen}, \bibinfo{person}{Paul Denny}, \bibinfo{person}{Stephen MacNeil}, \bibinfo{person}{Sami Sarsa}, \bibinfo{person}{Seth Bernstein}, \bibinfo{person}{Joanne Kim}, \bibinfo{person}{Andrew Tran}, {and} \bibinfo{person}{Arto Hellas}.} \bibinfo{year}{2023}\natexlab{a}.
\newblock \showarticletitle{Comparing Code Explanations Created by Students and Large Language Models}.
\newblock \bibinfo{journal}{\emph{arXiv preprint arXiv:2304.03938}} (\bibinfo{year}{2023}).
\newblock


\bibitem[Leinonen et~al\mbox{.}(2023b)]%
        {leinonen2023using}
\bibfield{author}{\bibinfo{person}{Juho Leinonen}, \bibinfo{person}{Arto Hellas}, \bibinfo{person}{Sami Sarsa}, \bibinfo{person}{Brent Reeves}, \bibinfo{person}{Paul Denny}, \bibinfo{person}{James Prather}, {and} \bibinfo{person}{Brett~A Becker}.} \bibinfo{year}{2023}\natexlab{b}.
\newblock \showarticletitle{Using large language models to enhance programming error messages}. In \bibinfo{booktitle}{\emph{Proceedings of the 54th ACM Technical Symposium on Computer Science Education V. 1}}. \bibinfo{pages}{563--569}.
\newblock


\bibitem[MacNeil et~al\mbox{.}(2023a)]%
        {macneil2024decoding}
\bibfield{author}{\bibinfo{person}{Stephen MacNeil}, \bibinfo{person}{Paul Denny}, \bibinfo{person}{Andrew Tran}, \bibinfo{person}{Juho Leinonen}, \bibinfo{person}{Seth Bernstein}, \bibinfo{person}{Arto Hellas}, \bibinfo{person}{Sami Sarsa}, {and} \bibinfo{person}{Joanne Kim}.} \bibinfo{year}{2023}\natexlab{a}.
\newblock \showarticletitle{Decoding Logic Errors: A Comparative Study on Bug Detection by Students and Large Language Models}.
\newblock \bibinfo{journal}{\emph{arXiv preprint arXiv:2311.16017}} (\bibinfo{year}{2023}).
\newblock


\bibitem[MacNeil et~al\mbox{.}(2023b)]%
        {macneil2023experiences}
\bibfield{author}{\bibinfo{person}{Stephen MacNeil}, \bibinfo{person}{Andrew Tran}, \bibinfo{person}{Arto Hellas}, \bibinfo{person}{Joanne Kim}, \bibinfo{person}{Sami Sarsa}, \bibinfo{person}{Paul Denny}, \bibinfo{person}{Seth Bernstein}, {and} \bibinfo{person}{Juho Leinonen}.} \bibinfo{year}{2023}\natexlab{b}.
\newblock \showarticletitle{Experiences from Using Code Explanations Generated by Large Language Models in a Web Software Development E-Book}. In \bibinfo{booktitle}{\emph{Proc. SIGCSE'23}}. \bibinfo{publisher}{ACM}, \bibinfo{numpages}{6}~pages.
\newblock


\bibitem[MacNeil et~al\mbox{.}(2022a)]%
        {macneil2022automatically}
\bibfield{author}{\bibinfo{person}{Stephen MacNeil}, \bibinfo{person}{Andrew Tran}, \bibinfo{person}{Juho Leinonen}, \bibinfo{person}{Paul Denny}, \bibinfo{person}{Joanne Kim}, \bibinfo{person}{Arto Hellas}, \bibinfo{person}{Seth Bernstein}, {and} \bibinfo{person}{Sami Sarsa}.} \bibinfo{year}{2022}\natexlab{a}.
\newblock \showarticletitle{Automatically Generating CS Learning Materials with Large Language Models}.
\newblock \bibinfo{journal}{\emph{arXiv preprint arXiv:2212.05113}} (\bibinfo{year}{2022}).
\newblock


\bibitem[MacNeil et~al\mbox{.}(2022b)]%
        {macneil2022generating}
\bibfield{author}{\bibinfo{person}{Stephen MacNeil}, \bibinfo{person}{Andrew Tran}, \bibinfo{person}{Dan Mogil}, \bibinfo{person}{Seth Bernstein}, \bibinfo{person}{Erin Ross}, {and} \bibinfo{person}{Ziheng Huang}.} \bibinfo{year}{2022}\natexlab{b}.
\newblock \showarticletitle{Generating Diverse Code Explanations Using the GPT-3 Large Language Model}. In \bibinfo{booktitle}{\emph{Proc. of the 2022 ACM Conf. on Int. Computing Education Research - Volume 2}}. \bibinfo{publisher}{ACM}, \bibinfo{pages}{37–39}.
\newblock
\showISBNx{9781450391955}


\bibitem[Madkins et~al\mbox{.}(2019)]%
        {madkins2019culturally}
\bibfield{author}{\bibinfo{person}{Tia~C Madkins}, \bibinfo{person}{Alexis Martin}, \bibinfo{person}{Jean Ryoo}, \bibinfo{person}{Kimberly~A Scott}, \bibinfo{person}{Joanna Goode}, \bibinfo{person}{Allison Scott}, {and} \bibinfo{person}{Frieda McAlear}.} \bibinfo{year}{2019}\natexlab{}.
\newblock \showarticletitle{Culturally relevant computer science pedagogy: From theory to practice}. In \bibinfo{booktitle}{\emph{2019 research on equity and sustained participation in engineering, computing, and technology (RESPECT)}}. IEEE, \bibinfo{pages}{1--4}.
\newblock


\bibitem[Morales-Chicas et~al\mbox{.}(2019)]%
        {morales2019computing}
\bibfield{author}{\bibinfo{person}{Jessica Morales-Chicas}, \bibinfo{person}{Mauricio Castillo}, \bibinfo{person}{Ireri Bernal}, \bibinfo{person}{Paloma Ramos}, {and} \bibinfo{person}{Bianca~L Guzman}.} \bibinfo{year}{2019}\natexlab{}.
\newblock \showarticletitle{Computing with relevance and purpose: A review of culturally relevant education in computing.}
\newblock \bibinfo{journal}{\emph{International Journal of Multicultural Education}} \bibinfo{volume}{21}, \bibinfo{number}{1} (\bibinfo{year}{2019}), \bibinfo{pages}{125--155}.
\newblock


\bibitem[Murphy et~al\mbox{.}(2012)]%
        {murphy2012ability}
\bibfield{author}{\bibinfo{person}{Laurie Murphy}, \bibinfo{person}{Sue Fitzgerald}, \bibinfo{person}{Raymond Lister}, {and} \bibinfo{person}{Ren\'{e}e McCauley}.} \bibinfo{year}{2012}\natexlab{}.
\newblock \showarticletitle{Ability to 'explain in Plain English' Linked to Proficiency in Computer-Based Programming}. In \bibinfo{booktitle}{\emph{Proc. of the Ninth Annual Int. Conf. on Int. Computing Education Research}}. \bibinfo{publisher}{ACM}, \bibinfo{pages}{111–118}.
\newblock
\showISBNx{9781450316040}


\bibitem[Prather et~al\mbox{.}(2023)]%
        {prather2023robots}
\bibfield{author}{\bibinfo{person}{James Prather}, \bibinfo{person}{Paul Denny}, \bibinfo{person}{Juho Leinonen}, \bibinfo{person}{Brett~A Becker}, \bibinfo{person}{Ibrahim Albluwi}, \bibinfo{person}{Michelle Craig}, \bibinfo{person}{Hieke Keuning}, \bibinfo{person}{Natalie Kiesler}, \bibinfo{person}{Tobias Kohn}, \bibinfo{person}{Andrew Luxton-Reilly}, {et~al\mbox{.}}} \bibinfo{year}{2023}\natexlab{}.
\newblock \showarticletitle{The robots are here: Navigating the generative ai revolution in computing education}.
\newblock In \bibinfo{booktitle}{\emph{Proceedings of the 2023 Working Group Reports on Innovation and Technology in Computer Science Education}}. \bibinfo{pages}{108--159}.
\newblock


\bibitem[Robins et~al\mbox{.}(2003)]%
        {robins2003learning}
\bibfield{author}{\bibinfo{person}{Anthony Robins}, \bibinfo{person}{Janet Rountree}, {and} \bibinfo{person}{Nathan Rountree}.} \bibinfo{year}{2003}\natexlab{}.
\newblock \showarticletitle{Learning and teaching programming: A review and discussion}.
\newblock \bibinfo{journal}{\emph{Computer science education}} \bibinfo{volume}{13}, \bibinfo{number}{2} (\bibinfo{year}{2003}), \bibinfo{pages}{137--172}.
\newblock


\bibitem[San~Chee(1993)]%
        {san1993applying}
\bibfield{author}{\bibinfo{person}{Yam San~Chee}.} \bibinfo{year}{1993}\natexlab{}.
\newblock \showarticletitle{Applying Gentner's theory of analogy to the teaching of computer programming}.
\newblock \bibinfo{journal}{\emph{International journal of man-machine studies}} \bibinfo{volume}{38}, \bibinfo{number}{3} (\bibinfo{year}{1993}), \bibinfo{pages}{347--368}.
\newblock


\bibitem[Sanders and McCartney(2016)]%
        {sanders2016threshold}
\bibfield{author}{\bibinfo{person}{Kate Sanders} {and} \bibinfo{person}{Robert McCartney}.} \bibinfo{year}{2016}\natexlab{}.
\newblock \showarticletitle{Threshold concepts in computing: past, present, and future}. In \bibinfo{booktitle}{\emph{Proceedings of the 16th Koli Calling international conference on computing education research}}. \bibinfo{pages}{91--100}.
\newblock


\bibitem[Sarsa et~al\mbox{.}(2022)]%
        {sarsa2022automatic}
\bibfield{author}{\bibinfo{person}{Sami Sarsa}, \bibinfo{person}{Paul Denny}, \bibinfo{person}{Arto Hellas}, {and} \bibinfo{person}{Juho Leinonen}.} \bibinfo{year}{2022}\natexlab{}.
\newblock \showarticletitle{Automatic Generation of Programming Exercises and Code Explanations Using Large Language Models}. In \bibinfo{booktitle}{\emph{Proc. of the 2022 ACM Conf. on Int. Computing Education Research - Volume 1}}. \bibinfo{publisher}{ACM}, \bibinfo{pages}{27–43}.
\newblock
\showISBNx{9781450391948}


\bibitem[Seel(2017)]%
        {seel2017model}
\bibfield{author}{\bibinfo{person}{Norbert~M Seel}.} \bibinfo{year}{2017}\natexlab{}.
\newblock \showarticletitle{Model-based learning: A synthesis of theory and research}.
\newblock \bibinfo{journal}{\emph{Educational Technology Research and Development}}  \bibinfo{volume}{65} (\bibinfo{year}{2017}), \bibinfo{pages}{931--966}.
\newblock


\bibitem[Tran et~al\mbox{.}(2023)]%
        {tran2023using}
\bibfield{author}{\bibinfo{person}{Andrew Tran}, \bibinfo{person}{Linxuan Li}, \bibinfo{person}{Egi Rama}, \bibinfo{person}{Kenneth Angelikas}, {and} \bibinfo{person}{Stephen MacNeil}.} \bibinfo{year}{2023}\natexlab{}.
\newblock \showarticletitle{Using Large Language Models to Automatically Identify Programming Concepts in Code Snippets}. In \bibinfo{booktitle}{\emph{Proc. of the 2023 ACM Conf. on Int. Computing Education Research - Volume 2}}, Vol.~\bibinfo{volume}{1}. \bibinfo{publisher}{ACM}, \bibinfo{pages}{563--569}.
\newblock
\urldef\tempurl%
\url{https://doi.org/10.1145/3568812.3603482}
\showDOI{\tempurl}


\bibitem[Vieira et~al\mbox{.}(2017)]%
        {vieira2017writing}
\bibfield{author}{\bibinfo{person}{Camilo Vieira}, \bibinfo{person}{Alejandra~J Magana}, \bibinfo{person}{Michael~L Falk}, {and} \bibinfo{person}{R~Edwin Garcia}.} \bibinfo{year}{2017}\natexlab{}.
\newblock \showarticletitle{Writing in-code comments to self-explain in computational science and engineering education}.
\newblock \bibinfo{journal}{\emph{ACM Transactions on Computing Education}} \bibinfo{volume}{17}, \bibinfo{number}{4} (\bibinfo{year}{2017}).
\newblock


\bibitem[Vihavainen et~al\mbox{.}(2015)]%
        {vihavainen2015benefits}
\bibfield{author}{\bibinfo{person}{Arto Vihavainen}, \bibinfo{person}{Craig~S Miller}, {and} \bibinfo{person}{Amber Settle}.} \bibinfo{year}{2015}\natexlab{}.
\newblock \showarticletitle{Benefits of self-explanation in introductory programming}. In \bibinfo{booktitle}{\emph{Proc. of the 46th ACM Technical Symposium on Computer Science Education}}. \bibinfo{pages}{284--289}.
\newblock


\bibitem[Whalley et~al\mbox{.}(2006)]%
        {whalley2006australasian}
\bibfield{author}{\bibinfo{person}{Jacqueline~L. Whalley}, \bibinfo{person}{Raymond Lister}, \bibinfo{person}{Errol Thompson}, \bibinfo{person}{Tony Clear}, \bibinfo{person}{Phil Robbins}, \bibinfo{person}{P.~K.~Ajith Kumar}, {and} \bibinfo{person}{Christine Prasad}.} \bibinfo{year}{2006}\natexlab{}.
\newblock \showarticletitle{An Australasian Study of Reading and Comprehension Skills in Novice Programmers, Using the Bloom and SOLO Taxonomies}. In \bibinfo{booktitle}{\emph{Proc. of the 8th Australasian Conf. on Computing Education - Volume 52}}. \bibinfo{publisher}{Australian Computer Society, Inc.}, \bibinfo{address}{AUS}, \bibinfo{pages}{243–252}.
\newblock
\showISBNx{1920682341}


\bibitem[Zastudil et~al\mbox{.}(2023)]%
        {cindyGenerating}
\bibfield{author}{\bibinfo{person}{Cynthia Zastudil}, \bibinfo{person}{Magdalena Rogalska}, \bibinfo{person}{Christine Kapp}, \bibinfo{person}{Jennifer Vaughn}, {and} \bibinfo{person}{Stephen MacNeil}.} \bibinfo{year}{2023}\natexlab{}.
\newblock \showarticletitle{Generative AI in Computing Education: Perspectives of Students and Instructors}. In \bibinfo{booktitle}{\emph{2023 IEEE Frontiers in Education Conference (FIE)}}. \bibinfo{pages}{1--9}.
\newblock
\urldef\tempurl%
\url{https://doi.org/10.1109/FIE58773.2023.10343467}
\showDOI{\tempurl}


\end{thebibliography}


\end{document}